\documentstyle[psfig]{caosp}
\begin{document}
\pubyear{1998}
\volume{27}

\firstpage{341}
\title{Gallium abundances in mercury-manganese stars}
\author{M. M. Dworetsky \and C. M. Jomaron \and C. A. Smith}
\institute{University College London, Gower Street, London, UK
WC1E 6BT}
\maketitle

\hyphenation{a-mounts Bi-del-man Hg-Mn Cor-liss}
\newcommand{\logg}{\mbox{$\log g$}}

\begin{abstract}

There is a widespread assertion in the literature that the optical Ga
lines give much higher abundances than the UV lines.  We have determined
Ga abundances in HgMn stars taking the observed hyperfine structure of the
optical Ga {\sc{ii}} lines into account.  This reduces these abundances to
within 0.2 dex of the values from the resonance lines.


\end{abstract}

\section{Gallium in optical spectra of HgMn stars}

Ga {\sc{ii}} was first identified in HgMn and He-weak stars by Bidelman
(1962), who identified four lines in the 4250 - 4260 \AA\ region. The
recent studies of gallium in HgMn stars by Keith Smith (1995, 1996)  found
high overabundances from the UV resonance lines of Ga~{\sc{ii}} and
Ga~{\sc{iii}} (between 2 and 4 orders of magnitude).  Most curve-of-growth
analyses based on the optical lines gave much higher abundances than the
UV lines.  This difference was greatest for the strongest lines and ranged
up to 1 dex in several cases (see Smith 1995, 1996; Takada-Hidai et al. 
1986).  Several explanations have been suggested (extreme stratification,
NLTE line formation, etc).  We have tested the hypothesis of Smith that
hyperfine structure in the lines might be able to account for this.  We
used simplified line models based on the laboratory spectroscopy of
Bidelman \& Corliss (1962) and a theoretical model by Lanz et al. (1993),
with {\it gf}-values from Ryabchikova \& Smirnov (1994).  Our observations
were Lick Hamilton \'{E}chelle spectra taken using the CAT in 1994-1997
and AAT service observations.  We observed nearly the entire Smith \&
Dworetsky (1993)  sample of HgMn stars.  We used the LTE spectrum
synthesis code {\sc{uclsyn}}, developed over many years at UCL by MMD and
KCS.  The code allows us to adjust input abundances, and simulate
instrumental profiles, rotation, and binarity. 

The ability to allow for contaminating blends is critically important
for Ga~{\sc{ii}}.  This is shown in the calculations for $\kappa$ Cnc
in Fig. 1 for four blue lines and one red line.  The tick marks for Ga
are at the positions of the main observed (or computed) hfs
components, with lengths roughly proportional to the strengths of the
components.  Other tick marks indicate the positions of the most
important blends.  It is immediately apparent that the most severe
blends are in $\lambda$4262 (Cr~{\sc{ii}}) and $\lambda$6334 
(Ne~{\sc{i}}) and if these are not taken into account one can seriously
overestimate the abundance of Ga.

Our results are given in Table 1 and show that the mean difference
in abundance is only 0.2 dex when hfs is taken into account along with
blends.  There is no trend with strength of the lines evident in our
results.  Much of the difference may be due to the crude model used
for hfs. Better laboratory observations are needed to extend this
work.  Meanwhile we will be checking further using theoretical hfs
predictions.

\begin{figure}
\begin{center} 
\leavevmode 
\psfig{figure=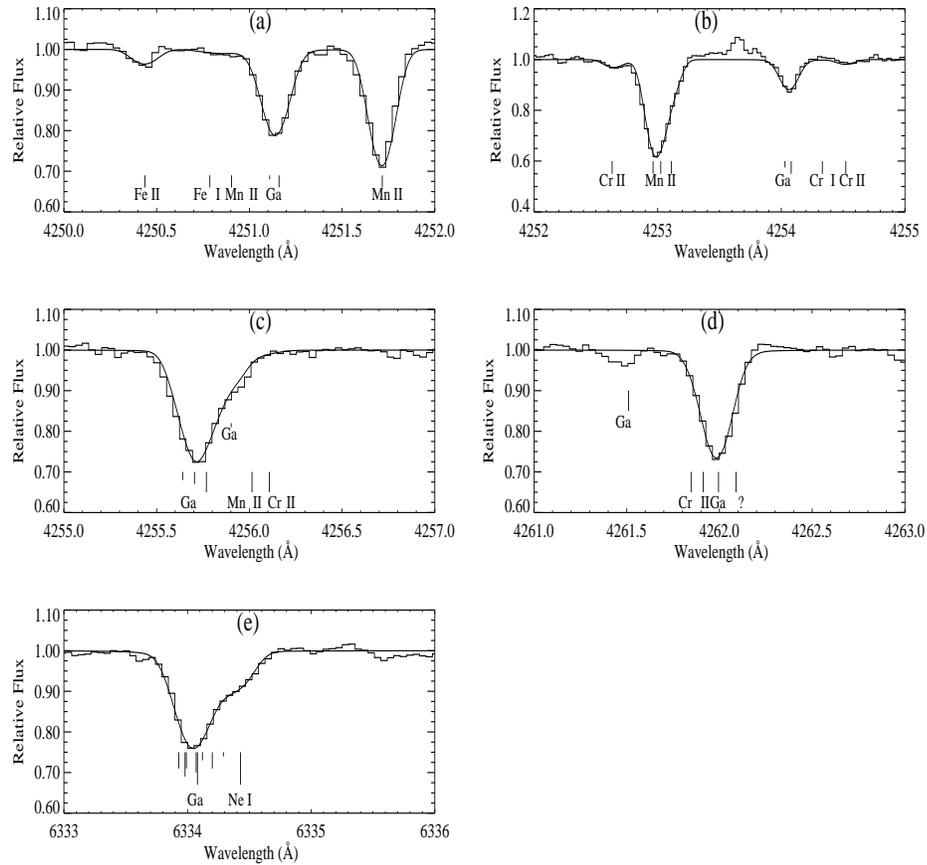,width=125mm,height=120mm}
\caption{(a) - (e): Observed (histograms) and synthetic (continuous
lines) spectra of the Ga {\sc{ii}} lines $\lambda\lambda$4251, 4254,
4255, 4262, and 6334 in the HgMn star $\kappa$ Cancri.}
\label{fig2}
\end{center} 
\end{figure}

\begin{table}
\caption{Gallium abundances}
\label{table:abundances}
\begin{tabular}{lccclcccc}
\hline

\small

Star &
\multicolumn{5}{c}{Ga\ {\sc{ii}} Abundances (log N(H) = 12)} &
Mean &$\sigma$ &Pub.\\
     &$\lambda$4251 &$\lambda$4254 &$\lambda$4255 &$\lambda$4262
&$\lambda$6334 & & &UV$^a$\\
\hline
{87 Psc} &5.60 &$\leq$6.0 &5.55 &5.80$^b$ &5.65 &5.65 &0.11 &5.45\\
{53 Tau} &5.55 &$\leq$5.7 &5.60 &5.40$^b$ &5.82 &5.59 &0.17 &5.65\\
{$\mu$ Lep} &7.15 &7.00 &6.75 &7.00$^b$ &6.75 &6.93 &0.18 &6.50\\
{HR 1800} &$\leq$4.8 &$\leq$4.8 &$\leq$4.8 &$\leq$4.8 &$\leq$5.0 
&$\leq$4.8 &0.09 &4.80\\
{33 Gem} &5.60 &$\leq$5.0 &$\sim$5.3 &$\leq$5.5 &$\leq$5.0 &$\leq$5.3 
&0.28 &5.20\\
{HR 2676} &$\leq$4.3 &$\leq$5.0 &$\leq$4.3 &$\leq$4.7 &$\leq$4.0 
&$\leq$4.5 &0.39 &4.00\\
{HR 2844} &7.18 &7.15 &6.95 &7.00$^b$ &6.85 &7.03 &0.14 &6.75\\
{$\kappa$ Cnc} &7.00 &7.00 &6.83 &6.70 &6.72 &6.85 &0.15 &6.60\\
{36 Lyn} &$\leq$6.3 &$\leq$6.5 &$\leq$5.8 &$\leq$5.7 &5.80 &$\leq$6.0 
&0.36 &5.10\\
{$\upsilon$ Her} &6.43 &6.32 &6.14 &6.25$^b$ &6.20 &6.27 &0.11 &6.05\\
{$\phi$ Her} &6.22 &$\sim$6.3 &5.75 &6.10$^b$ &5.82 &6.04 &0.24 
&5.70\\  
{28Her} &$\leq$4.8 &$\leq$6.1 &$\leq$4.7 &$\leq$5.2 &$\leq$5.2 
&$\leq$5.2 &0.55 &4.75\\
{HR 6997} &6.82 &7.00 &6.68 &6.50 &6.40 &6.68 &0.24 &6.45\\
{112 Her} &6.62 &6.62 &6.36 &6.38 &6.25 &6.45 &0.17 &6.35\\
{HR 7143} &6.75 &6.81 &6.52 &6.60$^b$ &6.51 &6.64 &0.14 &6.35\\   
{HR 7361} &6.90 &6.87 &6.66 &6.70 &6.58 &6.74 &0.14 &6.35\\
{46 Aql} &$\leq$4.4 &$\leq$5.5 &$\leq$4.5 &$\leq$4.6 &$\leq$4.4 
&$\leq$4.7 &0.47 &3.85\\
{HR 7664} &5.85 &5.90 &5.70 &5.55 &5.60 &5.72 &0.15 &5.60\\
{HR 7775} &6.66 &6.73 &6.38 &6.65$^b$ &5.90 &6.46 &0.34 &6.35\\
{$\beta$ Scl} &6.55 &6.70 &6.35 &6.35$^b$ &--- &6.49 
&0.17 &6.25\\
$\overline{\log (A/A_{\rm vis})}$ &+0.12 &+0.18 &$-0.09$ &$-0.04$ &$-0.15$ 
&--- &--- &$-0.22$\\  
\hline
\end{tabular}

$^a$Smith (1996); $^b${Ga} {\sc{ii}} $\lambda$4262 badly blended with 
{Cr} {\sc{ii}} line.\\ \end{table}

\acknowledgements
Research on chemically peculiar stars at UCL is supported by PPARC
grant GR/K58500 and travel to telescopes by PPARC grant GR/K60107.
Support from The Nuffield Foundation (NUF-URB97) and the Frank Norman
Jenkins Fund is gratefully acknowledged for C. A. Smith.

\end{document}